%% file: Arxiv_main.tex
\documentclass{article}

\usepackage[top=2.5cm,bottom=2.5cm,right=2cm,left=2cm]{geometry}

\usepackage{stfloats}
\usepackage{amsmath}
\usepackage{bbm}
\usepackage{amssymb}
\usepackage{enumitem}
\usepackage{amsthm}
\usepackage{graphicx}
\usepackage[version=4]{mhchem}
\usepackage{hyperref}
\usepackage{xcolor} 
\usepackage{authblk}
\usepackage{caption}
\usepackage{setspace}% make it double space
\usepackage{cleveref}
\usepackage{siunitx}
\usepackage{soul}
\usepackage[acronym]{glossaries}
\makeglossaries

\usepackage{xr} % for \externaldocument
\makeatletter
\newcommand*{\addFileDependency}[1]{% argument=file name and extension
  \typeout{(#1)}
  \@addtofilelist{#1}
  \IfFileExists{#1}{}{\typeout{No file #1.}}
}
\makeatother
\newcommand*{\myexternaldocument}[2][SI-]{%
    \externaldocument[#1]{#2}%
    \addFileDependency{#2.tex}%
    \addFileDependency{#2.aux}%
}
\myexternaldocument[SI-]{SI_main}
\usepackage{color}
\hypersetup{
        colorlinks=true,
        citecolor=blue,
        urlcolor  = black,
        linkcolor = blue
}

\DeclareCaptionType{extfig}[Extended Data Figure][Extended Data Figure]

\DeclareCaptionLabelSeparator{colon}{:\ }
\crefname{extfig}{Extended Data Figure}{Extended Data Figure}
\crefname{subextfig}{Extended Data Figure}{Extended Data Figure}
\input{glossary}
\title{Life cycle assessment for all organic chemicals}

\begin{document}
%\linenumbers
\author[1,2]{Shaohan Chen}
\author[1]{Tim Langhorst}
\author[1]{Julian Nöhl}
\author[2,3]{Christopher Oberschelp}
\author[1]{Martin Pillich}
\author[1,*]{Johannes Schilling} 
\author[1,2,*]{André Bardow}

\affil[1]{Laboratory of Energy and Process Systems Engineering (EPSE), ETH Zurich, 8092 Zurich, Switzerland}
\affil[2]{NCCR Catalysis, Switzerland}
\affil[3]{Chair of Ecological Systems Design (ESD), ETH Zurich, 8092 Zurich, Switzerland}

\affil[*]{To whom correspondence should be addressed. E-mail: \texttt{abardow@ethz.ch} and \texttt{jschilling@ethz.ch}}

\maketitle

\begin{abstract}
Chemicals are embedded in nearly every aspect of modern society, yet their production poses substantial sustainability concerns. Achieving a sustainable chemical industry requires detailed \acrfull{lca}; however, current assessments face many unknowns due to limited, partly inconsistent, and untransparent data coverage since existing \acrfull{lci} databases account for only a tiny fraction of traded chemicals. Here, we introduce the \acrfull{crystal} framework, which automatically generates consistent and transparent \acrshort{lci} data for organic chemicals based on their molecular structure using retrosynthesis and machine-learned gate-to-gate inventories. Using the predictive power of \acrshort{crystal}, we create a consistent database for more than \num{70000} organic chemicals, comprising over \num{110000} transparent \acrshort{lci} datasets that quantify both feedstock and energy demands, together with associated auxiliary materials, biosphere flows, and waste flows. From this comprehensive database, we identify \num{50} key environmental hotspots driving high impacts of organic chemical production across multiple environmental categories and pivotal hub chemicals that are most critical for downstream chemical production. In providing this comprehensive data foundation, the \acrshort{crystal} framework offers systematic guidance for targeted engineering and policy interventions. Its transparent, modular nature is designed to shift chemical \acrshort{lca} from a reliance on ``unknown unknowns'' to a collaboratively improvable mapping of ``known unknowns''.
\end{abstract}

\section{Introduction}\label{sec:introduction}
Chemicals and their derivatives are contained in \SI{95}{\percent} of manufactured goods in our society \cite{ICCA_Oxford2019}. Their production is the largest industrial consumer of fossil resources, the third largest global emitter of \ce{CO2} \cite{IEAreport}, and a major contributor to air and water pollution with serious risks to human health \cite{IEAreport, woodruff2024health}. Achieving global climate targets, therefore, requires a rapid and robust transition of the chemical industry towards sustainability. The progress of this transition depends on prioritizing environmental hotspots -- processes and products with the highest potential for impact reduction -- and replacing chemicals of concern, as promoted by the European Commission’s \textit{Safe and Sustainable by Design} initiative \cite{JRC, van2019safe}. However, these environmental hotspots remain largely unknown. 

The challenge is scale and complexity: \numrange{40000}{60000} chemicals are traded globally \cite{UNEP_ICCA_knowledge_sharing}, linked through intricate value chains. Reducing the resulting impacts effectively requires integrated system analysis connecting the potential improvements of upstream value chains with downstream production impacts. Such an integrated system analysis demands comprehensive, transparent, and consistent quantification of environmental impacts \cite{international2006environmental} based on detailed \acrfull{lci} data (i.e., the corresponding mass and energy flows) for all chemical production processes. Yet, such \acrshort{lci} data is unknown for most of the chemicals traded today and exists only for fewer than 2,000 chemicals in established \acrshort{lca} databases \cite{ecoinvent2025, carbonminds_database} (see \Cref{Fig:framework}a). Expanding the databases requires time-consuming, manual data collection for each missing chemical, making integrated system analysis infeasible at scale. This bottleneck hampers the ability to focus research and development on the chemical production processes that reduce environmental impacts, slowing the transition to a sustainable chemical industry. 

Expanding \acrshort{lca} coverage to the full breadth of the chemical sector has inspired the use of machine learning to predict the environmental impacts of chemical production \cite{zhang2024enhanced, kleinekorte2023appropriate, baxevanidis2022group, Gao2026GNN}. These approaches promise scale but are constrained by the scarcity of \acrshort{lca} training data and often operate as black boxes, offering limited interpretability of the underlying reaction pathway or \acrshort{lci} data. Efforts such as SemaNet \cite{chen2025semanet} address data gaps by inferring missing \acrshort{lci} flows from textual process descriptions, yet they do not explicitly capture the underlying reaction pathway for chemical production. First-principles approaches, such as stoichiometry-based methods \cite{langhorst2023stoichiometry}, provide a more transparent alternative but require detailed knowledge of the underlying reaction pathways -- knowledge that remains labor-intensive and time-consuming to extract from the literature.

Here, we present \acrfull{crystal}, a fully predictive, pathway-resolved \acrshort{lca} framework for organic chemical production that requires only the molecular structure of the target chemical as input. \acrshort{crystal} automatically generates transparent and consistent \acrshort{lci} data and pinpoints the environmentally optimal production pathways, enabling large-scale environmental impact assessments for organic chemicals. Using \acrshort{crystal}, we map a global chemical reaction network, identify key environmental hotspots across multiple impact categories relative to climate change, and highlight crucial processes whose improvement could deliver the greatest sustainability gains.

\section{The \texorpdfstring{\acrshort{crystal}}{CRYSTAL} framework: Automatic life cycle inventory generation for organic chemicals}

\begin{figure}[htbp]
	\begin{minipage}[t] {\textwidth}
		\centerline{\includegraphics[width=1\textwidth]{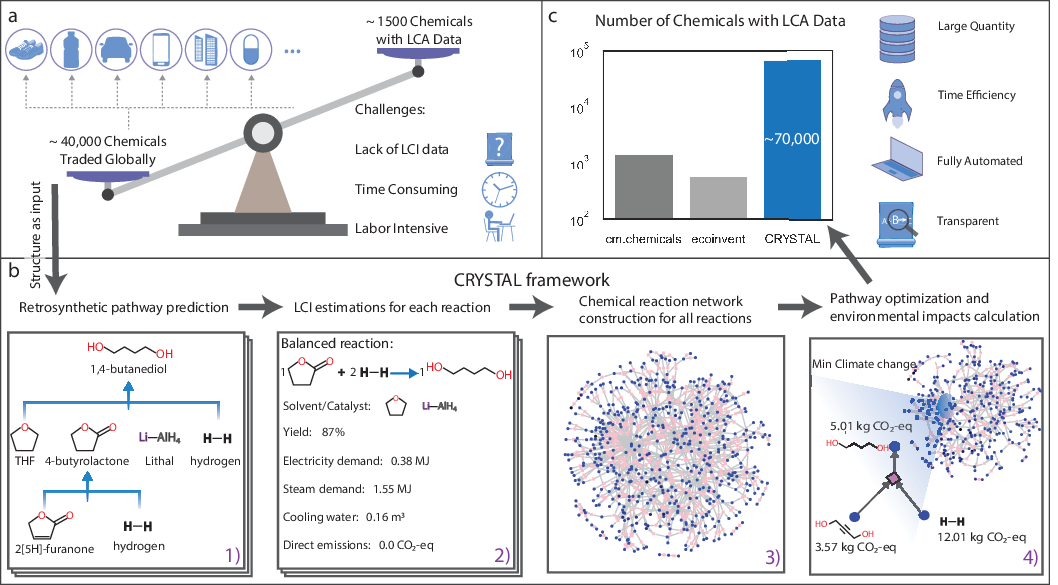}}
	\end{minipage}
\caption{\small\textbf{The \acrshort{crystal} framework automatically and transparently predicts cradle-to-gate \acrshort{lci} data for organic chemical production and identifies environmentally optimal production pathways.} \acrshort{crystal} balances efficient \acrshort{lci} data generation with the need to assess the environmental impacts of a vast number of traded chemicals, closing critical data gaps in life cycle assessments. Based on the molecular structure of the target chemical, we 1) predict retrosynthesis pathways, 2) estimate the \acrshort{lci} data, i.e., material demands, energy demands, waste streams, and direct emissions for each reaction step, 3) construct a \acrfull{crn} to transfer information across reactions, and 4) optimize the \acrshort{crn} based on the \acrfull{lcia} scores calculated from the predicted \acrshort{lci} data to identify environmentally optimal production pathways using user-specified environmental objectives. The framework flexibly adapts to various \acrshort{lcia} methods and background databases. 
} \label{Fig:framework}
\end{figure}

The \acrshort{crystal} framework (\Cref{Fig:framework}b) integrates domain knowledge from chemistry, engineering, and optimization to systematically and consistently quantify optimized cradle-to-gate environmental impacts of organic chemicals (details described in the Supplementary Text). For each new target chemical, \acrshort{crystal} commences by generating feasible synthesis routes through retrosynthesis powered by machine learning \cite{schwaller2020predicting}, which provides the reactants and auxiliaries for a chemical reaction. These synthesis routes are recursively generated back-propagating through the value chain until all precursors are available in the underlying \acrshort{lca} database (e.g., ecoinvent \cite{wernet2016ecoinvent,frischknecht2005ecoinvent,ecoinvent2025} or cm.chemicals \cite{carbonminds_database,stellner2023_carbonminds}). Since inorganic compounds are occasionally required as reactants or auxiliary materials to produce organic chemicals, we incorporate literature-reported inorganic reaction pathways for critical inorganic precursors not present in the underlying \acrshort{lca} database to complement the retrosynthesis tool. Subsequently, the identified reaction equations are balanced and byproducts are identified via an optimization algorithm to estimate feedstock demands and amounts of byproducts as well as waste streams. Next, the reaction yields, energy demands, and auxiliary material demands are predicted using decision trees trained on industrial process data \cite{langhorst2025reaction}. The treatment of waste and byproducts is guided by a hierarchical rule-based model that integrates best practices from industry (details described in the Supplementary Text) and waste incineration models \cite{doka2003life, doka2013updates,meys2021achieving}. Thereby, all required \acrshort{lci} data are obtained and the corresponding \acrfull{lcia} scores of the target chemical can be calculated. The modular structure of \acrshort{crystal} enables an efficient exchange of each module (e.g., the underlying \acrshort{lca} database or approach for energy demand estimation).

Since the retrosynthesis tool was not designed to predict environmentally optimal reaction pathways, and improved routes may arise via shared intermediates, we combined all identified reactions into a unified \acrfull{crn}. The \acrshort{crn} is optimized using a graph-theory-based algorithm to propagate the \acrshort{lci} information and identify environmentally preferable pathways. This \acrshort{crn} can be flexibly extended by the inclusion of industrially relevant reactions if required. 

Following this comprehensive approach, \acrshort{crystal} automatically identifies optimal, transparent production pathways jointly with their \acrshort{lci} data and quantifies the corresponding environmental impacts consistently. 
We apply \acrshort{crystal} in a high-throughput screening of market-relevant chemicals identified from multiple regulatory and industrial sources, including the \acrshort{reach} database \cite{REACH_2025}, the \acrshort{dippr} database \cite{wilding1998dippr}, the \acrshort{oecd} database \cite{OECD_HPV_chemicals}, the U.S. Environmental Protection Agency \acrshort{cdr} databases \cite{CDR2020}, the Scandinavian Norden \acrshort{spin} database \cite{SPIN2025} and the Japanese \acrshort{meti} database \cite{Meti2025} (details described in the Supplementary Text).
This screening generates consistent \acrshort{lci} data and environmentally optimized production pathways for approximately \num{70000} target chemicals and their intermediates in a matter of hours. The resulting \acrshort{lca} database is \num{40} times larger than the largest existing \acrshort{lca} databases \cite{ecoinvent2025,carbonminds_database} (\Cref{Fig:framework}c), and can be continuously expanded using \acrshort{crystal} to address critical \acrshort{lca} data gaps and support future research.

\section{Validation of \texorpdfstring{\acrshort{crystal}}{CRYSTAL} based on state-of-the-art \texorpdfstring{\acrshort{lca}}{LCA} databases}

\begin{figure}[htbp]
	\begin{minipage}[t] {\textwidth}
		\centerline{\includegraphics[width=1\textwidth]{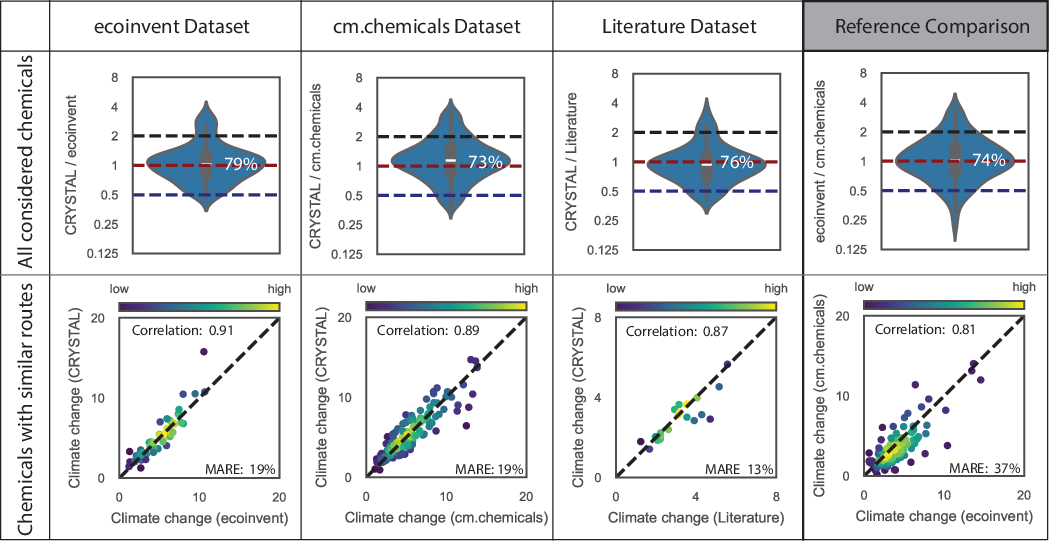}}
	\end{minipage}
	\caption{\small\textbf{Performance of \acrshort{crystal} on three state-of-the-art \acrshort{lca} databases for climate change impact.} The first three columns present the validation results of the \acrshort{crystal} framework on the ecoinvent \cite{wernet2016ecoinvent,frischknecht2005ecoinvent,ecoinvent2025} (\num{257} organic chemicals (version 3.10, cut-off) assessed with ReCiPe 2016 v1.03, midpoint (H)), cm.chemicals \cite{carbonminds_database,stellner2023_carbonminds} (\num{746} organic chemicals based on IPCC 2021 global warming potential (GWP100)), and a literature database \cite{langhorst2023stoichiometry} (\num{143} organic chemicals assessed with ecoinvent, version 3.5, cut-off) based on ReCiPe v1.13, midpoint (H)), using the corresponding database as the basis and a \acrlong{loo} approach. The ``Reference Comparison'' column compares the ecoinvent and cm.chemicals databases for \num{232} chemicals identified as the complete intersection of the two databases with available \acrshort{smiles} representations, serving as a benchmark for evaluating \acrshort{crystal}'s performance. The violin plots of the first three columns summarize \acrshort{crystal}’s overall performance on each database, indicating the percentage of \textit{climate change} (unit: \unit{\kilogram} \ce{CO2}-eq per \unit{\kilogram} of the target chemical) predictions falling within an acceptable accuracy range, based on predicted-to-reference value ratios. Dashed lines indicate the acceptable range defined by the \acrfull{aace} accuracy bounds (\SIrange{50}{200}{\percent} of the environmental impact defined in the corresponding state-of-the-art database). Violin plots display the distribution between the 5th and 95th percentiles to highlight this range. The parity plots of the first three columns illustrate \acrshort{crystal}’s Pearson correlation and \acrfull{mare} when the last reaction step of the predicted routes is the same as the one of the reference databases.} \label{Fig:validation}
\end{figure}

Unlike existing machine-learning approaches that predict only aggregated environmental impacts of chemicals, \acrshort{crystal} predicts the underlying \acrshort{lci} data in a transparent manner, which then serves as the basis for calculating the corresponding environmental impacts. As such \acrshort{lca} data barely exists, a particular challenge arises in validating a method like \acrshort{crystal}.
We validate the performance of \acrshort{crystal} by comparing its predicted environmental impacts with reference values from three state-of-the-art \acrshort{lca} databases \cite{ecoinvent2025,carbonminds_database,langhorst2023stoichiometry}, illustrated here for climate change impacts (\Cref{Fig:validation}). To ensure consistent comparison, we use each state-of-the-art database as the underlying reference for \acrshort{crystal} and applied a \acrfull{loo} approach (details described in the Supplementary Text), harmonizing the background processes and model assumptions across the datasets.

More than \SI{73}{\percent} of \acrshort{crystal}'s predictions fall within a factor of \num{2} of the state-of-the-art reference impacts, a range defined as acceptable in cost engineering standards by the Association for the Advancement of Cost Engineering International (Class 5: \SIrange{50}{200}{\percent}) \cite{dysert2016aace}. Our results demonstrate that \acrshort{crystal}'s deviations are on par with the variability observed between the existing state-of-the-art \acrshort{lca} databases themselves (column entitled ``Reference Comparison'' in \Cref{Fig:validation}). Such discrepancies between databases largely stem from differences in upstream background processes and model assumptions (e.g., market mixes of global chemical production) \cite{cullen2024reducing}. Notably, the considered cm.chemicals database version \cite{carbonminds_database,stellner2023_carbonminds} adopts ecoinvent as its background database. Comparisons between fully independent \acrshort{lca} databases are thus expected to exhibit even larger discrepancies.

One source for the remaining deviations between \acrshort{crystal} and the reference database is a different pathway to synthesize the same chemical. Because different production pathways are employed in practice, such differences cannot \textit{a priori} be regarded as errors. 
When the last predicted reaction step aligns with the reference pathway, \acrshort{crystal} demonstrates high predictive accuracy, achieving a Pearson correlation coefficient of \num{0.9} and \acrlong{mare} of \SI{19}{\percent}, with \SI{96}{\percent} and \SI{100}{\percent} of the predictions falling within the acceptable range for ecoinvent \cite{ecoinvent2025,wernet2016ecoinvent,frischknecht2005ecoinvent} and cm.chemicals \cite{carbonminds_database,stellner2023_carbonminds}, respectively (\Cref{Fig:validation}). \acrshort{crystal} predictions are closest to ecoinvent \cite{ecoinvent2025,wernet2016ecoinvent,frischknecht2005ecoinvent} and cm.chemicals \cite{carbonminds_database} databases when a single reaction step is predicted, with \SI{81}{\percent} and \SI{79}{\percent} of predictions within the acceptable range, respectively (Figures S37 and S38). Deviations between \acrshort{crystal}'s predictions and reference values increase with the number of predicted reaction steps, due to accumulating discrepancies between the predicted and reference pathways. In the \acrshort{loo} approach, the need to predict more than one reaction step typically indicates a discrepancy between the predicted and reference pathways, since all precursors should, in principle, be available in the underlying \acrshort{lca} database.

\acrshort{crystal} generalizes beyond climate change impact, as demonstrated by the validation across all environmental impact categories (Figures S28-S36). In particular, \SI{87}{\percent} of \acrshort{crystal}'s predictions for \textit{Energy resources: non-renewable, fossil} fall within the defined acceptable range, achieving a Pearson correlation coefficient of \num{0.89} when the last reaction step of the predicted pathway matches the reference pathway.

To further validate the performance on more structurally and functionally complex chemicals, such as pharmaceuticals, we benchmark \acrshort{crystal} against two published databases \cite{parvatker2019cradle, huber2022approach} and an industrial dataset provided by an industrial partner (Figures S39, S41, and S43). \acrshort{crystal} successfully reproduces both the overall trends and relative magnitude of the environmental impacts in the reference data when predicting the same last reaction step, achieving Pearson's correlation coefficients of \num{0.95} and \num{0.98} (logarithmic values used to account for scale differences across chemicals) on two published databases \cite{parvatker2019cradle, huber2022approach}. For the industrial dataset, \acrshort{crystal} attains a Pearson correlation coefficient of \num{0.81} in overall performance, despite the absence of detailed information on reaction pathways due to confidentiality.

While \acrshort{lca} lacks a definitive ground truth for validation \cite{heijungs2024probability}, our results demonstrate strong alignment between \acrshort{crystal} and established benchmarks. We acknowledge that \acrshort{crystal} provides an optimistic estimate of environmental impacts, as the framework systematically identifies the most favorable chemical routes currently modeled. Nevertheless, by extending the reach of \acrshort{lca} to tens of thousands of chemicals, \acrshort{crystal} serves as a transparent starting point to close critical data gaps, incentivize the sharing of primary industrial data, and strengthen \acrshort{lca} studies across the chemical industry.

\section{Environmental hotspots in organic chemical production beyond climate change} \label{sec:hotspots}

\begin{figure}[htbp]
	\begin{minipage}[t] {\textwidth}
		\centerline{\includegraphics[width=1\textwidth]{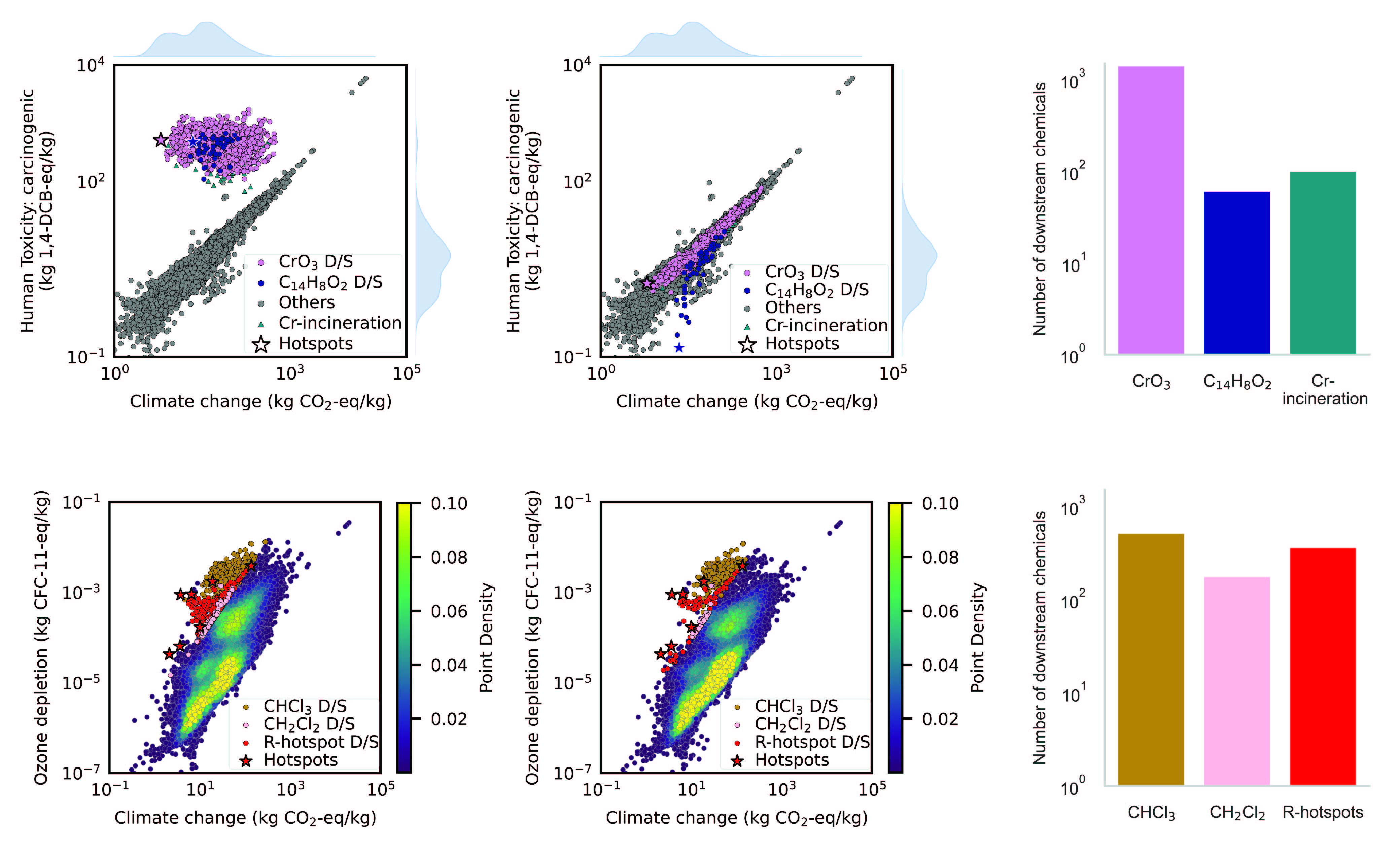}}
	\end{minipage}
	\caption{\small\textbf{Pinpointing environmental hotspots across impact categories relative to climate change to support chemical manufacturers, regulatory authorities, and \acrshort{lca} data providers.} 
Top row: \textit{Climate change} versus \textit{human toxicity: carcinogenic} for pathways optimized for \textit{climate change} impact. Chromium trioxide and anthraquinone, together with chromium-containing waste incineration, are highlighted as relative environmental hotspots in \textit{human toxicity: carcinogenic}, along with their downstream chemicals (left). Regulating chromium(VI) emissions for these two hotspot reactants and for the chromium-containing waste treatment effectively reduces \textit{human toxicity: carcinogenic} of all downstream chemicals (middle). Number of downstream chemicals (D/S) associated with these hotspots (right).
Bottom row: \textit{Climate change} versus \textit{ozone depletion} for pathways optimized for (left) \textit{climate change} and (middle) \textit{ozone depletion} impact. Relative environmental hotspots in climate-change-optimized pathways are highlighted together with their downstream chemicals (i.e., five upstream reactants (R-hotspots) and two solvents, chloroform and dichloromethane). To emphasize changes in point density, only downstream chemicals with disproportionately high ozone depletion impacts are highlighted (middle). Number of downstream chemicals associated with these hotspots (right). ecoinvent version 3.10 (cut-off) \cite{wernet2016ecoinvent,frischknecht2005ecoinvent,ecoinvent2025} and \acrshort{lcia} method ReCiPe 2016 v1.03, midpoint (H) are used as the underlying \acrshort{lca} database.}  \label{Fig:hotspots}
\end{figure}

\acrshort{crystal} enables the systematic identification of environmental hotspots (details described in the Supplementary Text) along the chemical production value chain, including raw feedstocks, auxiliary materials, waste treatment, and energy demand. Because current assessment practice is largely focused on climate change \cite{life_cycle_assessment_chemical_industry_review}, we identify relative environmental hotspots that disproportionately contribute to other environmental impacts. 
Of the approximately \num{70000} chemicals in the \acrshort{crystal} database, \num{1616} exhibit disproportionately high human toxicity (carcinogenic) in their climate-change-optimized pathways (\Cref{Fig:hotspots} top row).
Tracing the full production value chain of these chemicals reveals two upstream reactants as the dominant contributors (details described in the Supplementary Text): chromium trioxide (\acrshort{casrn} 1333-82-0, \acrshort{lcia} data from ecoinvent -- \num{1451} downstream products), anthraquinone (\acrshort{casrn} 84-65-1, \acrshort{lcia} data from ecoinvent -- \num{62} downstream products). The remaining chemicals with high human toxicity (carcinogenic) primarily arise from chromium-containing waste treatment.

The elevated human toxicity (carcinogenic) associated with the hotspots chromium trioxide and anthraquinone originates from both their production and the treatment of (chromium-containing) waste in upstream processes, as reported in ecoinvent \cite{althaus2007life}. The waste treatment model implemented in \acrshort{crystal} (details described in the Supplementary Text) assumes that non-valuable byproducts (including chromium-containing waste) are incinerated \cite{rada2021regulatory, EuroEnvironmental2024, astrup2005chromium, secretariat1997technical} following the Doka model \cite{doka2003life,doka2013updates}, resulting in further emissions of Chromium(VI) to groundwater, rivers, and the atmosphere \cite{doka2003life}. 
However, increasing regulatory pressure \cite{BaselConvention2025,blasenbauer2020legal} is driving the chemical industry to reduce emissions of Chromium(VI). An effective mitigation strategy is to convert Chromium(VI) into the less toxic and more stable Chromium(III) \cite{ATSDR2012Chromium}. In our simulation, converting Chromium(VI) to Chromium(III) within the production value chains of chromium trioxide and anthraquinone (details described in the Supplementary Text), as well as during the treatment of chromium-containing wastes, effectively eliminates the disproportionately high human carcinogenic toxicity observed for these hotspots and their downstream chemicals (\Cref{Fig:hotspots}, top row, middle). Our results highlight the importance of establishing Chromium(VI) reduction measures as an industrial standard in waste incineration and landfilling and underscore the urgent need to ensure that current regulatory practices are incorporated into \acrshort{lci} databases and waste treatment models \cite{doka2003life,doka2013updates}.

These results reveal that targeted regulation and process improvements at upstream stages of chemical production identified as relative environmental hotspots can substantially reduce human toxicity impacts. The results for other environmental impact categories are reported in Figures S46-S53.
More broadly, our findings emphasize the need to integrate up-to-date industrial practices and regulatory advances into \acrshort{lca} inventories and models, including the current \acrshort{crystal} framework, to ensure accurate environmental impact assessments and to provide robust guidance for sustainable chemical design and policy.

For the trade-off between ozone depletion and climate change, we identify five clusters of chemicals along their climate-change-optimized pathways (\Cref{Fig:hotspots} -- bottom row, left) using a Gaussian mixture clustering algorithm (see \Cref{Extend:data1}a and details described in the Supplementary Text).
Among chemicals with disproportionately high ozone-depletion impacts, we identify seven upstream contributors as the primary drivers of ozone depletion (\Cref{Fig:hotspots} -- bottom, left; methods in the Supplementary Text): two solvents, chloroform (\acrshort{casrn} 67-66-3, data from ecoinvent -- \num{525} downstream products) and dichloromethane (\acrshort{casrn} 75-09-2, data from ecoinvent -- \num{181} downstream products); and five reactants (all data from ecoinvent -- in total \num{368} downstream products). The high ozone-depletion impacts of these contributors stem from direct emissions of halogenated alkanes or nitrous oxide during production, or from the use of chloroform as a precursor.

By shifting the objective from minimizing climate change impact to ozone depletion, \num{27600} chemicals adopt alternative production pathways, mitigating the ozone-depletion impacts associated with climate-change-optimized pathways. For these chemicals, ozone-depletion impacts can be substantially reduced by \SI{50}{\percent} with only a \SI{19}{\percent} increase in climate change impacts (\Cref{Fig:hotspots} -- bottom, middle). Ozone-depletion impacts decrease across all five clustering centroids, with the largest reduction observed for the cluster exhibiting the highest initial impact (see \Cref{Extend:data1}b). 
Nevertheless, for some chemicals, mitigation of ozone-depletion impacts remains constrained by the reactions available within the \acrshort{crn}.

Our analysis reinforces the objectives of the Montreal Protocol \cite{ozone_depletion_report} by demonstrating the potential environmental benefits of reducing or substituting chloroform and dichloromethane as solvents to mitigate ozone depletion. Stringent control of direct emissions of halogenated alkanes and nitrous oxide remains essential, as indicated by the five identified upstream hotspot reactants. As the retrosynthesis tool \cite{schwaller2020predicting} continues to evolve through updated training data reflecting ongoing advances in the chemical industry, and the generated \acrshort{lci} data is expanded accordingly, the \acrshort{crystal} framework will support this activity by pinpointing actionable interventions in chemical production.

\section{Which chemicals matter most for improving sustainability?} \label{sec:hubs}

\begin{figure}[htbp]
	\begin{minipage}[t] {\textwidth}
		\centerline{\includegraphics[width=1\textwidth]{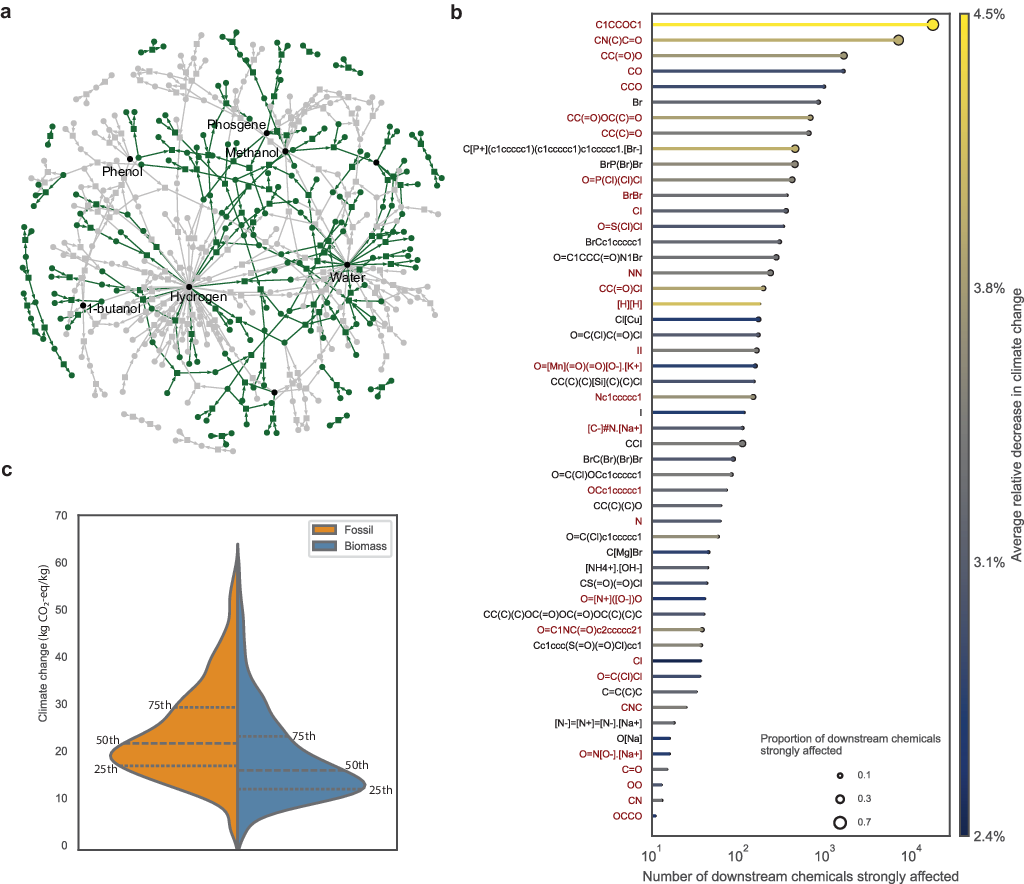}}
	\end{minipage}
	\caption{ \small\textbf{Hub chemicals identified to guide prioritization of process optimization strategies for enhancing the sustainability of the chemical industry.} 
    \textbf{a}) A representative example illustrating the structurally important role of a hub chemical within an illustrative chemical reaction network. The green part indicates the joint-optimal reactions across all environmental impact categories.
    \textbf{b}) Ranking of \num{52} hub chemicals by the number of strongly affected downstream chemicals, measured as those whose climate change impacts decrease by more than \SI{2}{\percent} in response to a \SI{10}{\percent} reduction in the impact of each hub chemical (details described in the Supplementary Text). Chemicals available in the underlying ecoinvent database are highlighted in red. 
    \textbf{c}) Switching from fossil- to bio-based ethanol pathways substantially reduces the climate change impacts for over \num{2000} strongly affected downstream products.} \label{Fig:map}
\end{figure}

Based on its underlying chemical reaction network, \acrshort{crystal} uncovers hub chemicals -- key intermediates whose environmental impacts influence large portions of the \acrshort{crn}. Because of their central role in connectivity and material flow (see \Cref{Fig:map}a), improving the sustainability of these hub chemicals can propagate benefits across the entire chemical industry \cite{jacob2018statistics}. Here, we define hub chemicals as those that (i) appear in at least \num{216} environmentally optimized downstream pathways (determined by the elbow method, see \Cref{Extend:data2}) and (ii) cause measurable propagation effects; i.e., a \SI{10}{\percent} decrease in their environmental impact translates into more than \SI{2}{\percent} lower impacts for more than \num{10} downstream chemicals. We refer to these downstream chemicals as \textit{strongly affected} nodes.

We identify \num{52} hub chemicals that exert broad, system-wide influence across the climate-change-optimized chemical reaction network (\Cref{Fig:map}b; detailed information in Table S9). Among these hub chemicals, tetrahydrofuran (THF; \acrshort{casrn} 109-99-9) is particularly impactful, strongly impacting over \num{17500} downstream chemicals (\SI{68}{\percent} of its total downstream chemicals) with an average \SI{4.5}{\percent} decrease in climate change impacts. This influence propagates hierarchically through up to \num{18} downstream reaction steps (\Cref{Extend:data3}a), reflecting THF's dual role as both feedstock and solvent. The widespread use of THF as a reaction medium \cite{Greed2022TurningSolvents} further underscores its strategic importance for targeted process improvement to achieve system-wide climate benefits. 

A similar pattern is observed for dimethylformamide (DMF; \acrshort{casrn} 68-12-2), which strongly impacts \num{7000} downstream chemicals and an average of \SI{4.0}{\percent} decrease in climate change impacts, again reflecting its dual function as feedstock and solvent (\Cref{Extend:data3}b).

Although methanol and ethanol appear in over \num{25000} and \num{19900} downstream pathways, respectively, only \SI{5}{\percent} and \SI{6}{\percent} of downstream chemicals are strongly affected in terms of climate change (\Cref{Extend:data3} b, c). This limited propagation reflects the relatively low climate change impact of their own production pathways: biomass fermentation for ethanol and natural gas reforming for methanol. Within our \acrshort{crn}, bio-based ethanol reduces the climate change impacts for \num{2047} strongly affected downstream chemicals by an average of \SI{6}{\kilogram} \ce{CO2}-eq compared to its fossil-based production, highlighting the substantial climate benefits achievable by transitioning hub chemicals to bio-based production (\Cref{Fig:map}c). However, this shift also entails environmental trade-offs, notably increased land and water use impacts (Figures S49 and S51).

Hub chemicals that rank highly for climate change often exert substantial influence across other environmental impact categories (\Cref{Extend:data4}; details described in the Supplementary Text). However, this pattern is not guaranteed: for instance, nitric acid strongly impacts more than \num{2800} downstream chemicals in ozone depletion due to nitrous oxide emissions, but it has only a limited influence on other impact categories. These findings highlight the importance of considering multiple environmental dimensions beyond climate change, as some hub chemicals exhibit category-specific impacts that would be overlooked in climate-change-focused assessments. Comparing optimized \acrshort{crn}s across environmental impact categories reveals that ozone depletion and marine eutrophication correlate least strongly with climate change (see \Cref{Extend:data5}; and analysis in Supplementary Text) and should therefore be prioritized when analyzing environmental trade-offs in chemical production.

While the identified hub chemicals may not reflect all real-world impacts due to model limitations, the results demonstrate how focusing on hub chemicals with cascading downstream effects can inform system-wide sustainability improvements, particularly as industrial data availability increases and our model assumptions continue to be refined.

\section{Enabling a sustainable chemical industry}

\acrshort{crystal} represents a step change in assessing the environmental impacts of chemical production by transparently and consistently predicting \acrshort{lci} data from retrosynthetic pathways. By automatically providing \acrshort{lci}s for organic chemicals with consistent assumptions, transparent reaction pathways, and uniform details, \acrshort{crystal} closes a critical data gap in \acrshort{lca}. Thus, we believe that the provided database will be very useful to \acrshort{lca} practitioners. Still, acknowledging G.E.P. Box's famous statement that ``all models are wrong, but some are useful'', we regard transparency as the crucial feature of the \acrshort{crystal} model and database: if an expert knows that a reaction path, solvent or yield are different in practice, these components of the \acrshort{lci}s can be replaced. To maximize continuous refinement, \acrshort{crystal} is designed as an open, community-oriented platform (open-source code and versioned datasets) that invites contributions from researchers, industry, and data providers. Community curation and structured feedback loops will iteratively improve prediction quality.

While the identified environmentally optimized pathways may not reflect real-world economically optimized industrial practice, preventing their direct use in environmental reporting,
\acrshort{crystal} allows identifying opportunities to improve organic chemical sustainability: By pinpointing hub chemicals and hotspot processes for targeted optimization, \acrshort{crystal} offers actionable guidance for industry, regulators, and \acrshort{lca} data providers.

Since the retrosynthetic pathway prediction is based on available patent literature, \acrshort{crystal} could be further enhanced by incorporating pathways based on bio-based reactions \cite{zuiderveen2023potential, lee2019comprehensive} and plastic waste recycling \cite{rahimi2017chemical}, enabling substitution of fossil feedstocks in critical chemicals and cascading system-wide environmental benefits (\Cref{sec:hotspots} and \Cref{sec:hubs}). Incorporation of production volumes will enable sector-wide optimization to identify synergies in joint chemical production. Coupling \acrshort{crystal} with retrosynthesis planning could guide chemical synthesis \cite{bustillo2023rise} toward inherently greener routes, establishing a self-reinforcing cycle of sustainable chemical innovation.

As the pace of novel chemical discovery accelerates, \acrshort{crystal} can provide rapid, automated, and transparent sustainability assessments from the earliest stages. This capability empowers chemists and engineers to evaluate emerging pathways rapidly, ensuring that innovation advances without unintended environmental trade-offs and fostering a chemical industry that is both productive and sustainable. Crucially, by incorporating community feedback into model and data updates, \acrshort{crystal} is positioned not only as a tool but also as a collaborative ecosystem to continuously improve \acrshort{lci} predictions and their real-world relevance.

\section*{Data and materials availability}
All data supporting the findings of this study are available in the main text or in the supplementary materials, except for data derived from commercial databases and the commercial databases themselves, which cannot be redistributed. The entire \acrshort{crystal} framework and corresponding \acrshort{crystal} database will be made available as soon as possible. Correspondence and requests for materials should be addressed to A.B.

\section*{Supplementary materials}

The Supplementary Text, including the mentioned Figures S28-S39, Figure S41, Figure S43, Figures S46-S53, and Table S9, is not included in this arXiv preprint and will be made available in a later version.

%% Glossary
\renewcommand{\glossarypreamble}{\glsfindwidesttoplevelname[\currentglossary]}
\setglossarystyle{alttree}

%\clearpage \addcontentsline{toc}{section}{Acronyms} \printglossary[type=\acronymtype,title=Acronyms,nonumberlist]

%\clearpage \addcontentsline{toc}{section}{Symbols} \printglossary[type=main,title=Symbols,nonumberlist]

% Bibliography
\bibliographystyle{naturemag}
\bibliography{references}

\section*{Acknowledgments}
This publication was developed as part of NCCR Catalysis (grant numbers 180544 and 225147), a National Centre of Competence in Research funded by the Swiss National Science Foundation. S.C. also acknowledges funding from the USorb-DAC Project, supported by a grant from The Grantham Foundation for the Protection of the Environment to RMI’s climate tech accelerator program, Third Derivative. We acknowledge the support from Dr. Teodoro Laino and his team from IBM, who work on the RXN tool. We acknowledge the helpful comments and suggestions provided by Arne K\"{a}telh\"{o}n of Carbon Minds during the review of the manuscript. M.P. acknowledges funding by the European Union’s Horizon 2020 research and innovation programme as part of the project CIRCULAR FOAM under grant agreement No. 101036854. This work reflects only the authors' views. It does not represent the view of the European Commission and the Commission is not responsible for any use that may be made of the information it contains.

\section*{Author contributions}
S.C. contributed to study conceptualization, methodology development, data curation, and validation, and wrote the original draft of the manuscript. T.L. assisted in developing the methodology. J.N. assisted with the data analysis and reviewed the manuscript. C.O. contributed to validation as well as data curation and assisted in developing the methodology. M.P. assisted with methodology development, contributed to user-interface development, and reviewed the manuscript. J.S. contributed to study conceptualization and data publication, assisted with methodology development and in writing the manuscript, and supervised the project. A.B. contributed to study conceptualization, methodology development, assisted in writing the manuscript, supervised the project, and acquired funding. All authors contributed to discussions and finalizing the manuscript. 

\section*{Conflicts of interest}
The authors declare the following financial interests and personal relationships that could be considered as potential competing interests: A.B. has served on review committees for research and development at ExxonMobil and TotalEnergies, companies active in both oil and gas and chemical production. A.B. holds ownership interests in firms that provide services to industry, some of which may operate in the chemical industry. In particular, A.B. has ownership interests in Carbon Minds, a company that supplies \acrshort{lca} databases used, among others, in this work to validate the \acrshort{crystal} model. T.L. joined Carbon Minds as an employee after his contributions to this work were completed. The remaining authors declare no known competing financial interests or personal relationships that could have influenced the work reported in this paper.

\newpage

\begin{extfig}[htbp]
	\begin{minipage}[t] {\textwidth}
		\centerline{\includegraphics[width=1\textwidth]{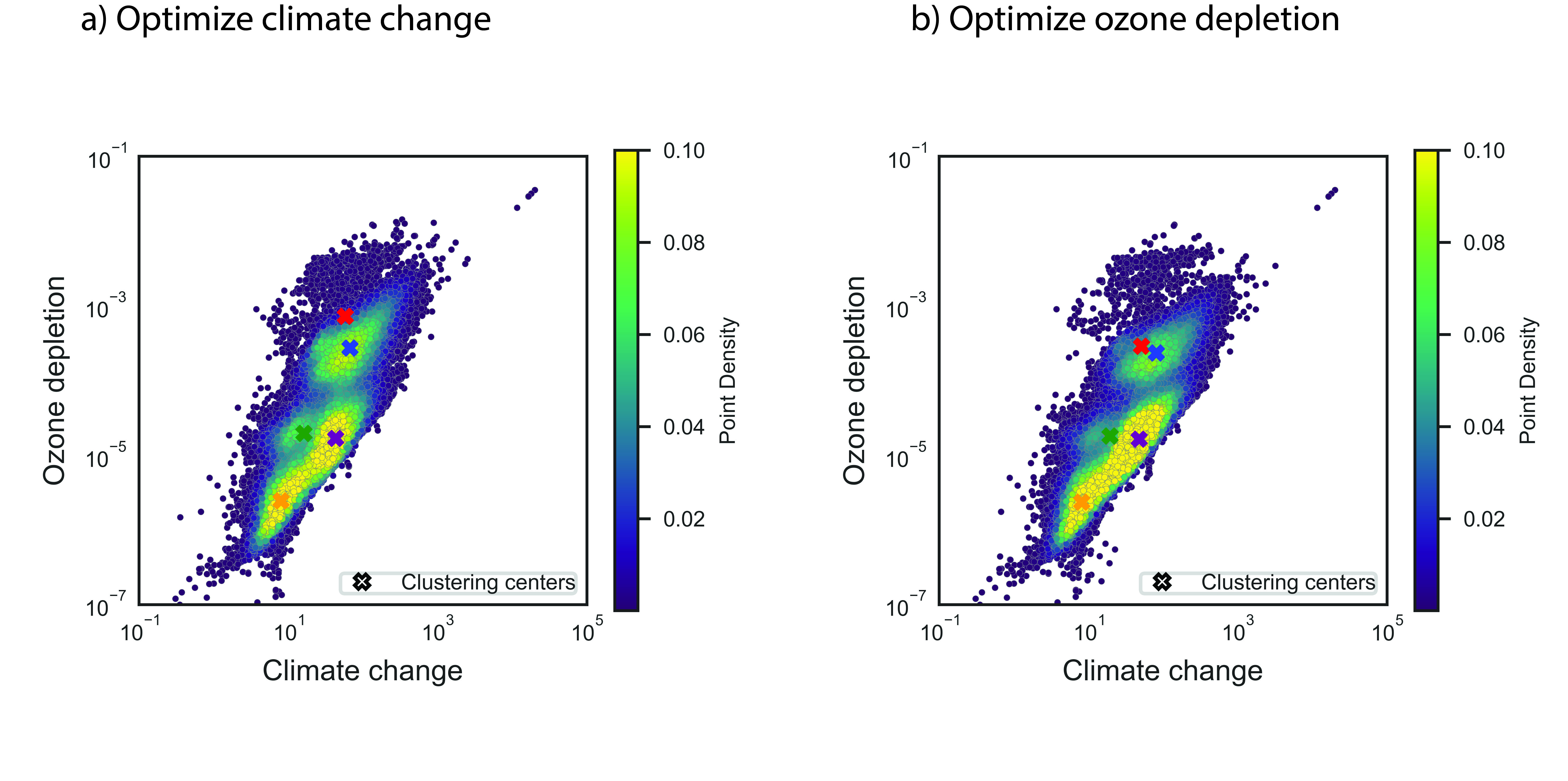}}
	\end{minipage}
	\caption{\textbf{a}) Clustering centroids of five clusters of chemicals along their climate-change-optimized pathways using a Gaussian mixture clustering algorithm. \textbf{b}) Clustering centroids of five clusters of chemicals along their ozone-depletion-optimized pathways -- details described in the Supplementary Text.} \label{Extend:data1}
\end{extfig}

\begin{extfig}[htbp]
	\begin{minipage}[t] {\textwidth}
		\centerline{\includegraphics{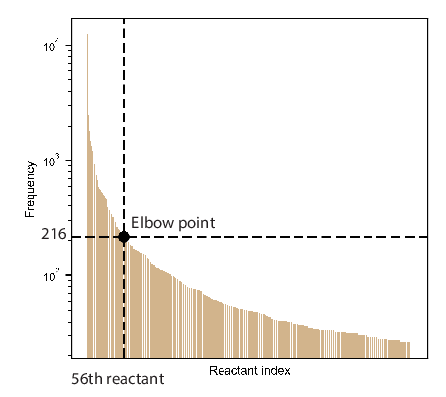}}
	\end{minipage}
	\caption{Chemicals most frequently used as reactants ($n=56$), identified using the Elbow method \cite{satopaa2011finding}.} \label{Extend:data2}
\end{extfig}

\newpage
\begin{extfig}[htbp]
	\begin{minipage}[t]{\textwidth}		\centerline{\includegraphics[width=1\textwidth]{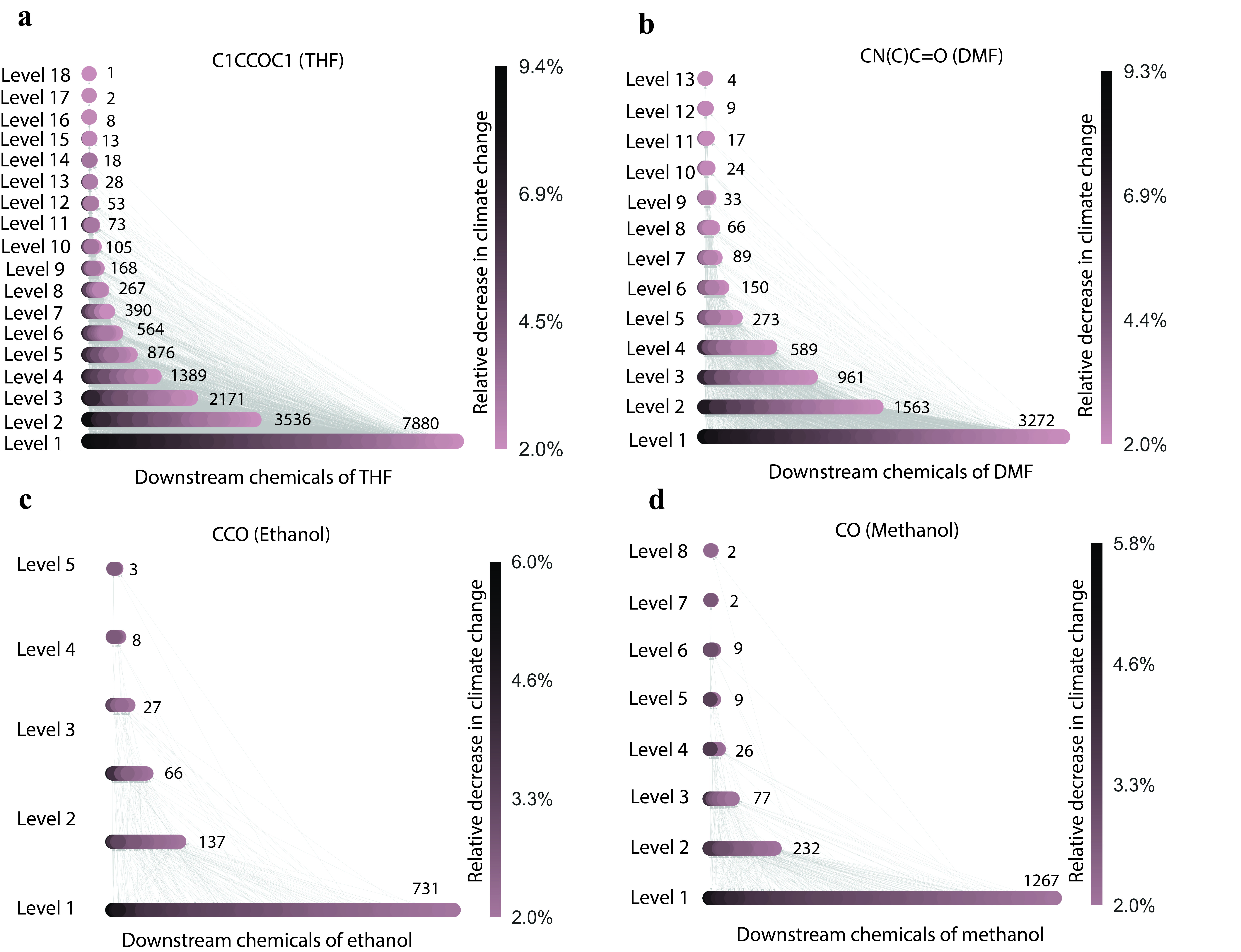}}
	\end{minipage}
	\caption{\small\textbf{Climate-change influence of hub chemicals on downstream products.} Relative reduction in downstream climate change impact resulting from a \SI{10}{\percent} reduction in the impact of each hub chemical: a) tetrahydrofuran (THF), b) dimethylformamide (DMF), c) ethanol, and d) methanol.} \label{Extend:data3}
\end{extfig}

\newpage

\begin{extfig}[htbp]
	\begin{minipage}[t] {\textwidth}
		\centerline{\includegraphics{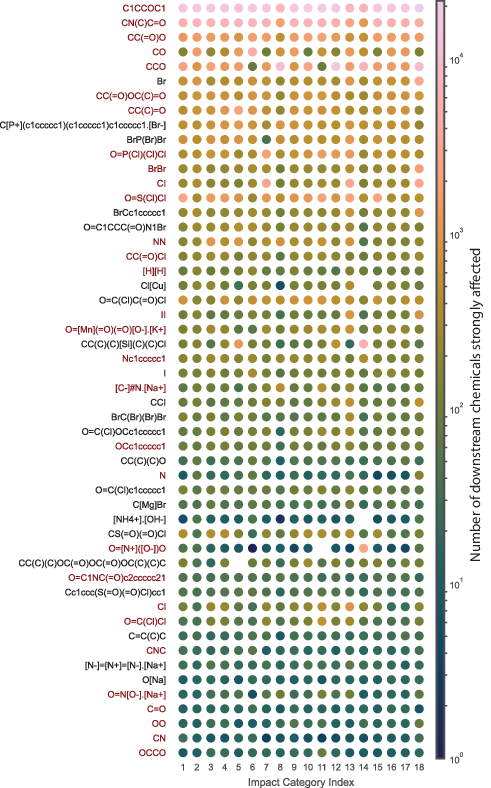}}
	\end{minipage}
	\caption{\small\textbf{Heat map showing the sensitivity of hub chemicals across multiple environmental impact categories.} Number of strongly affected downstream chemicals showing more than \SI{2}{\percent} impact decrease in response to a \SI{10}{\percent} decrease in each hub chemical, across environmental impact categories. The full names of the environmental impact categories are as follows: 1. Acidification: terrestrial, 2. Climate change, 3. Ecotoxicity: freshwater, 4. Ecotoxicity: marine, 5. Ecotoxicity: terrestrial, 6. Energy resources: non-renewable, fossil, 7. Eutrophication: freshwater, 8. Eutrophication: marine, 9. Human toxicity: carcinogenic, 10. Human toxicity: non-carcinogenic, 11. Ionising radiation, 12. Land use, 13. Material resources: metals/minerals, 14. Ozone depletion, 15. Particulate matter formation, 16. Photochemical oxidant formation: human health, 17. Photochemical oxidant formation: terrestrial ecosystems, 18. Water use.} \label{Extend:data4}
\end{extfig}
\newpage

\begin{extfig}[htbp]
	\begin{minipage}[t] {\textwidth}
		\centerline{\includegraphics[width=1\textwidth]{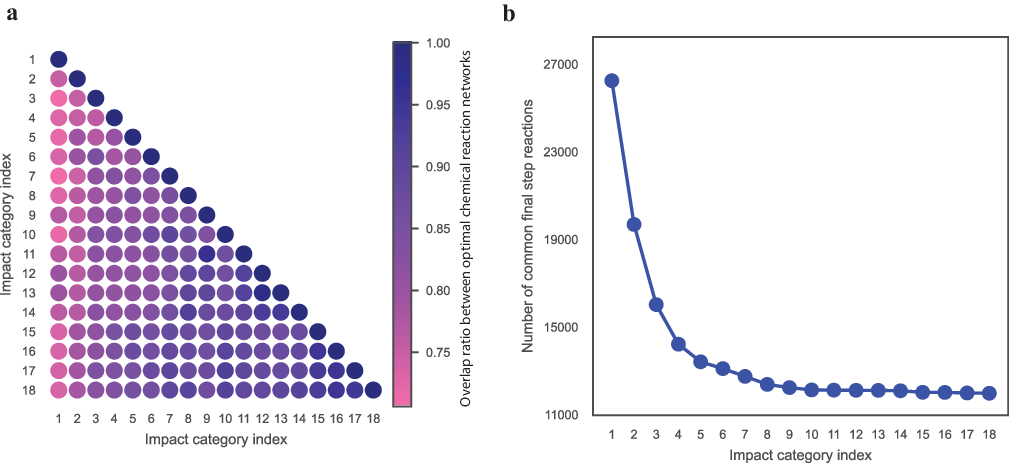}}
	\end{minipage}
	\caption{\small\textbf{Environmental impact categories influencing optimal pathway selection in the chemical reaction network.} \textbf{a}) Overlap ratios between chemical reaction networks optimized for \num{18} environmental impact categories. \textbf{b}) Decline in the number of common reactions among optimal networks as the number of environmental impact categories considered increases. For more than \num{11000} chemicals, the \acrshort{crn} optimization identifies a final reaction step that simultaneously minimizes all environmental impacts, whereas trade-offs across impact categories remain for the remaining chemicals.
    The full names of the environmental impact categories are as follows: 1. Ozone depletion, 2. Eutrophication: marine, 3. Water use, 4. Material resources: metals/minerals, 5. Ecotoxicity: terrestrial, 6. Land use, 7. Ionising radiation, 8.Energy resources: non-renewable, fossil, 9. Acidification: terrestrial, 10. Eutrophication: freshwater, 11. Particulate matter formation, 12. Photochemical oxidant formation: human health, 13. Photochemical oxidant formation: terrestrial ecosystems, 14. Climate change, 15. Ecotoxicity: freshwater, 16. Ecotoxicity: marine, 17. Human toxicity: non-carcinogenic, 18. Human toxicity: carcinogenic. Only chemicals with at least two reaction pathways in the network are included for rigorous evaluation.} \label{Extend:data5}
\end{extfig}
\end{document}

%% file: glossary.tex
\newacronym{aace}{AACE}{Association for the Advancement of Cost Engineering}

\newacronym{alop}{ALOP}{Agricultural Land Occupation Potential}

\newacronym{cactus}{CACTUS}{CADD Group Chemoinformatics Tools and User Services}

\newacronym{cas}{CAS}{Chemical Abstracts Service}

\newacronym{casrn}{CAS RN}{CAS Registry Number}

\newacronym{cdr}{CDR}{Chemical Data Reporting}

\newacronym{cml}{CML}{Centrum voor Milieukunde}

\newacronym{crn}{CRN}{Chemical Reaction Network}

\newacronym{crns}{CRNs}{chemical reaction networks}

\newacronym{dippr}{DIPPR}{Design Institute for Physical Properties}

\newacronym{eea}{EEA}{European Economic Area}

\newacronym{epa}{EPA}{Environmental Protection Agency}

\newacronym{fdp}{FDP}{Fossil Depletion Potential}

\newacronym{fetp}{FETPinf}{Freshwater Ecotoxicity Potential}

\newacronym{fep}{FEP}{Freshwater Eutrophication Potential}

\newacronym{ghg}{GHG}{Greenhouse Gas}

\newacronym{glo}{GLO}{Global (Generic global geography)}

\newacronym{gmm}{GMM}{Gaussian Mixture Model}

\newacronym{gwp}{GWP}{Global Warming Potential}

\newacronym{cc}{CC}{Climate Change Impact}

\newacronym{gwi}{GWI}{Global Warming Impact}

\newacronym{hpv}{HPV}{High Production Volume}

\newacronym{htp}{HTPinf}{Human Toxicity Potential}

\newacronym{irp}{IRP\_HE}{Ionising Radiation Potential}

\newacronym{ipcc}{IPCC}{Intergovernmental Panel on Climate Change}

\newacronym{idea}{IDEA}{Inventory Database for Environmental Analysis}

\newacronym{iur}{IUR}{Inventory Update Reporting}

\newacronym{kpi}{KPI}{Key Performance Indicator}

\newacronym{lasso}{LASSO}{Least Absolute Shrinkage and Selection Operator}

\newacronym{lca}{LCA}{Life Cycle Assessment}

\newacronym{lci}{LCI}{Life Cycle Inventory}

\newacronym{lcia}{LCIA}{Life Cycle Impact Assessment}

\newacronym{loo}{LOO}{Leave-One-Out}

\newacronym{mae}{MAE}{Mean Absolute Error}

\newacronym{medad}{MedAD}{Median Absolute Deviation}

\newacronym{mare}{MARE}{Mean Absolute Relative Error}

\newacronym{mdp}{MDP}{Metal Depletion Potential}

\newacronym{meti}{METI}{Ministry of Economy, Trade and Industry}

\newacronym{mep}{MEP}{Marine Eutrophication Potential}

\newacronym{metp}{METPinf}{Marine Ecotoxicity Potential}

\newacronym{mr:mm}{MR:MM}{Material Resources: Metals/Minerals}

\newacronym{ML}{ML}{Machine Learning}

\newacronym{nltp}{NLTP}{Natural Land Transformation Potential}

\newacronym{nn}{NN}{Neural Network}

\newacronym{nrtl}{NRTL}{nonrandom two-liquid model}

\newacronym{odp}{ODPinf}{Ozone Depletion Potential}

\newacronym{oecd}{OECD}{Organisation for Economic Co-operation and Development}

\newacronym{pep}{PEP}{Process Economics Programme}

\newacronym{pmfp}{PMFP}{Particulate Matter Formation Potential}

\newacronym{pmi}{PMI}{product mass intensity}

\newacronym{pofp}{POFP}{Photochemical Oxidant Formation Potential}

\newacronym{crystal}{CRYSTAL}{Chemical RetrosYnthesiS for Transparent Assessment of Life-cycles}

\newacronym{reach}{REACH}{Registration, Evaluation, Authorisation and Restriction of Chemicals Regulation}

\newacronym{rer}{RER}{Europe}

\newacronym{row}{RoW}{Rest-of-World}

\newacronym{rmse}{RMSE}{Root Mean Squared Error}

\newacronym{smiles}{SMILES}{Simplified molecular-input line-entry system}

\newacronym{spin}{SPIN}{Substances in Preparations in Nordic Countries}

\newacronym{spt}{SPT}{SMILES-to-Properties-Transformer}

\newacronym{tap}{TAP100}{Terrestrial Acidification Potential}

\newacronym{tetp}{TETPinf}{Terrestrial Ecotoxicity Potential}

\newacronym{TRL}{TRL}{Technological Readiness Level}

\newacronym{ulop}{ULOP}{Urban Land Occupation Potential}

\newacronym{wdp}{WDP}{Water Depletion Potential}

\newglossaryentry{alpha}{
    name=$\alpha$,
    description={tuning parameter (-)}}

\newglossaryentry{betaj}{
    name=$\beta_j$,
    description={penalty parameter for byproduct $j$ (-)}}

\newglossaryentry{BPminE}{
    name=$\mathrm{BPminE}$,
    description={minimum normal boiling point of reactants (\unit{\kelvin})}}

\newglossaryentry{BPmaxE}{
    name=$\mathrm{BPmaxE}$,
    description={maximum normal boiling point
of reactants (\unit{\kelvin})}}

\newglossaryentry{BPminP}{
    name=$\mathrm{BPminP}$,
    description={minimum normal boiling point of products (\unit{\kelvin})}}

\newglossaryentry{BPmaxP}{
    name=$\mathrm{BPmaxP}$,
    description={maximum normal boiling point
of products (\unit{\kelvin})}}

\newglossaryentry{B}{
    name=$\mathrm{B}$,
    description={Set of valuable byproducts (-)}}

\newglossaryentry{Cr}{
    name=$\mathrm{C_\mathrm{r}}$,
    description={Reactants coefficient of the reaction to produce the target product (-)}}

\newglossaryentry{D}{
    name=$D$,
    description={Set of all processes within the considered system boundaries to produce the target chemical (-)}}

\newglossaryentry{1ji}{
    name=$\boldsymbol{1}_{ji}$,
    description={binary variable, which indicates whether the $i$-th atom group appears in the $j$-th precursor (-)}}

\newglossaryentry{h_i}{
    name=$h_i$,
    description={Life Cycle Impact Assessment score of the production of chemical $i$ (unit dependent on the impact category and functional unit)}}

\newglossaryentry{lambda}{
    name=$\lambda$,
    description={allocation factor (-)}}

\newglossaryentry{p}{
    name=$p$,
    description={number of atom groups (-)}}

\newglossaryentry{m_crn}{
    name=$m$,
    description={number of reactions in the chemical reaction network (-)}}

\newglossaryentry{MP}{
    name=$M_\mathrm{P}$,
    description={molar weight of target product (\unit{\kilogram \per \kilo \mol})}}
    
\newglossaryentry{MBi}{
    name=$M_{\mathrm{B},i}$,
    description={molar weight of byproduct $i$ (\unit{\kilogram \per \kilo \mol})}}

\newglossaryentry{Mi}{
    name=$M_i$,
    description={molar weight of reactant $i$ (\unit{\kilogram \per \kilo \mol})}}

\newglossaryentry{mi}{
    name=$m_{\mathrm{R,}i}$,
    description={mass demand of reactant $i$ (\unit{\kilogram})}}

\newglossaryentry{mBi}{
    name=$m_{\mathrm{B,}i}$,
    description={produced mass of byproduct $i$ (\unit{\kilogram})}}

\newglossaryentry{mP}{
    name=$m_\mathrm{P}$,
    description={produced mass of the target product (\unit{\kilogram})}}

\newglossaryentry{mB}{
    name=$m_\mathrm{B}$,
    description={total mass of the byproducts (\unit{\kilogram})}}

\newglossaryentry{mER}{
    name=$m_\mathrm{R,excess}$,
    description={total mass of the unreacted (excess) reactants (\unit{\kilogram})}}

\newglossaryentry{mReag}{
    name=$m_\mathrm{reagents}$,
    description={total mass of the all reagents (\unit{\kilogram})}}

\newglossaryentry{q}{
    name=$q$,
    description={number of atom types (-)}}

\newglossaryentry{n_crn}{
    name=$n$,
    description={number of species in the chemical reaction network (-)}}

\newglossaryentry{nc}{
    name=$n_\mathrm{C}$,
    description={number of \ce{C} atoms (-)}}

\newglossaryentry{no}{
    name=$n_\mathrm{O}$,
    description={number of \ce{O} atoms (-)}}

\newglossaryentry{nh}{
    name=$n_\mathrm{H}$,
    description={number of \ce{H} atoms (-)}}

\newglossaryentry{n_AB}{
    name=$n_\mathrm{Atoms,B}$,
    description={number of atoms in the byproduct (-)}}

\newglossaryentry{omega}{
    name=$\omega$,
    description={acentric factor (-)}}

\newglossaryentry{p_i}{
    name=$p_i$,
    description={Life cycle inventory amount of process $i$ (unit dependent on the process).}
}

\newglossaryentry{pc}{
    name=$p_\mathrm{c}$,
    description={critical pressure (\unit{\bar})}}

\newglossaryentry{psat}{
    name=$p_\mathrm{sat}$,
    description={saturation pressure (\unit{\bar})}}

\newglossaryentry{R}{
    name=$\mathrm{R}$,
    description={Set of reactants (-)}}

\newglossaryentry{R2}{
    name=$\mathrm{R}^2$,
    description={coefficient of determination (-)}}

\newglossaryentry{ros}{
    name=$r_\mathrm{os}$,
    description={offset from trend line to identify hotspots in the environmental impacts (-)}}

\newglossaryentry{T}{
    name=$T$,
    description={temperature (\unit{\kelvin})}}

\newglossaryentry{Tb}{
    name=$T_\mathrm{b}$,
    description={normal boiling temperature (\unit{\kelvin})}}

\newglossaryentry{Tc}{
    name=$T_\mathrm{c}$,
    description={critical temperature (\unit{\kelvin})}}

\newglossaryentry{xi}{
    name=$x_i$,
    description={stoichiometric coefficient of reactant $i$ (-)}}

\newglossaryentry{yi}{
    name=$y_i$,
    description={stoichiometric coefficient of product $i$ (-)}}

\newglossaryentry{yP}{
    name=$y_\mathrm{P}$,
    description={stoichiometric coefficient of the target product (-)}}

%% file: references.bib
@article{baxevanidis2022group,
  author       = {Baxevanidis, Pantelis and Papadokonstantakis, Stavros and Kokossis, Antonis and Marcoulaki, Effie},
  title        = {Group Contribution-Based {LCA} Models to Enable Screening for Environmentally Benign Novel Chemicals in {CAMD} Applications},
  journal      = {AIChE Journal},
  year         = {2022},
  volume       = {68},
  number       = {3},
  pages        = {e17544},
  publisher    = {Wiley},
  doi          = {10.1002/aic.17544},
  url          = {https://doi.org/10.1002/aic.17544}
}

@article{kleinekorte2023appropriate,
  author       = {Kleinekorte, Johanna and Kleppich, Jonas and Fleitmann, Lorenz and Beckert, Verena and Blodau, Luise and Bardow, Andr{\'e}},
  title        = {{APPROPRIATE Life Cycle Assessment: A PROcess-Specific, PRedictive Impact AssessmenT Method for Emerging Chemical Processes}},
  journal      = {ACS Sustainable Chemistry \& Engineering},
  year         = {2023},
  volume       = {11},
  number       = {25},
  pages        = {9303--9319},
  publisher    = {American Chemical Society},
  doi          = {10.1021/acssuschemeng.2c07682},
  url          = {https://doi.org/10.1021/acssuschemeng.2c07682}
}

@article{langhorst2023stoichiometry,
  author       = {Langhorst, Tim and Winter, Benedikt and Roskosch, Dennis and Bardow, Andr{\'e}},
  title        = {Stoichiometry-Based Estimation of Climate Impacts of Emerging Chemical Processes: Method Benchmarking and Recommendations},
  journal      = {ACS Sustainable Chemistry \& Engineering},
  year         = {2023},
  volume       = {11},
  number       = {17},
  pages        = {6600--6609},
  publisher    = {American Chemical Society},
  doi          = {10.1021/acssuschemeng.2c07624},
  url          = {https://doi.org/10.1021/acssuschemeng.2c07624}
}

@article{schwaller2020predicting,
  author    = {Philippe Schwaller and Riccardo Petraglia and Valerio Zullo and Vishnu H. Nair and Rico Andreas Haeuselmann and Riccardo Pisoni and Costas Bekas and Anna Iuliano and Teodoro Laino},
  title     = {Predicting retrosynthetic pathways using transformer-based models and a hyper-graph exploration strategy},
  journal   = {Chemical Science},
  year      = {2020},
  volume    = {11},
  number    = {12},
  pages     = {3316--3325},
  publisher = {Royal Society of Chemistry},
  doi       = {10.1039/C9SC05704H},
  url       = {https://doi.org/10.1039/C9SC05704H}
}

@article{wernet2016ecoinvent,
  title={The ecoinvent Database Version 3 ({Part I}): Overview and Methodology},
  author={Wernet, Gregor and Bauer, Christian and Steubing, Bernhard and Reinhard, J{\"u}rgen and Moreno-Ruiz, Emilia and Weidema, Bo},
  journal={The International Journal of Life Cycle Assessment},
  volume={21},
  number = {9},
  pages={1218--1230},
  year={2016},
  publisher={Springer},
  doi = {10.1007/s11367-016-1087-8},
  url = {https://doi.org/10.1007/s11367-016-1087-8}
}

@article{frischknecht2005ecoinvent,
  title={The ecoinvent Database: Overview and Methodological Framework},
  author={Frischknecht, Rolf and Jungbluth, Niels and Althaus, Hans-J{\"o}rg and Doka, Gabor and Dones, Roberto and Heck, Thomas and Hellweg, Stefanie and Hischier, Roland and Nemecek, Thomas and Rebitzer, Gerald and others},
  journal={The International Journal of Life Cycle Assessment},
  volume={10},
  number = {1},
  pages={3--9},
  year={2005},
  publisher={Springer},
  doi = {10.1065/lca2004.10.181.1},
  url = {https://doi.org/10.1065/lca2004.10.181.1}
}

@article{langhorst2025reaction,
  author       = {Langhorst, Tim and Winter, Benedikt and Tuchschmid, Moritz and Roskosch, Dennis and Bardow, Andr{\'e}},
  title        = {From Reaction Stoichiometry to Life Cycle Assessment: Decision Tree-Based Estimation Tool},
  journal      = {ACS Environmental Au},
  year         = {2025},
  volume       = {},
  number       = {},
  pages        = {},
  publisher    = {American Chemical Society},
  doi          = {10.1021/acsenvironau.4c00065},
  url          = {https://doi.org/10.1021/acsenvironau.4c00065}
}

@article{wilding1998dippr,
  author       = {Wilding, W. Vincent and Rowley, Richard L. and Oscarson, John L.},
  title        = {DIPPR{\textregistered} Project 801 Evaluated Process Design Data},
  journal      = {Fluid Phase Equilibria},
  year         = {1998},
  volume       = {150},
  pages        = {413--420},
  publisher    = {Elsevier},
  doi          = {10.1016/S0378-3812(98)00341-0},
  url          = {https://doi.org/10.1016/S0378-3812(98)00341-0}
}

@article{woodruff2024health,
  author       = {Tracey J. Woodruff},
  title        = {Health Effects of Fossil Fuel--Derived Endocrine Disruptors},
  journal      = {The New England Journal of Medicine},
  year         = {2024},
  volume       = {390},
  number       = {10},
  pages        = {922--933},
  doi          = {10.1056/NEJMra2300476},
  url          = {https://doi.org/10.1056/NEJMra2300476}
}

@techreport{IEAreport,
  author       = {Araceli Fernandez Pales and Peter Levi},
  title        = {The Future of Petrochemicals: Towards More Sustainable Plastics and Fertilisers},
  institution  = {International Energy Agency (IEA)},
  year         = {2018},
  doi          = {10.1787/9789264307414-en},
  url          = {https://www.iea.org/reports/the-future-of-petrochemicals},
  publisher    = {OECD Publishing / International Energy Agency},
  address      = {Paris, France},
  lastaccessed = {2025-10-17}
}

@article{cullen2024reducing,
  author       = {Cullen, Luke and Meng, Fanran and Lupton, Rick and Cullen, Jonathan M.},
  title        = {Reducing uncertainties in greenhouse gas emissions from chemical production},
  journal      = {Nature Chemical Engineering},
  year         = {2024},
  volume       = {1},
  number       = {4},
  pages        = {311--322},
  publisher    = {Nature Publishing Group},
  doi          = {10.1038/s44286-024-00047-z},
  url          = {https://www.nature.com/articles/s44286-024-00047-z}
}

@misc{ecoinvent2025,
  author       = {{ecoinvent Association}},
  title        = {ecoinvent Database Version 3},
  year         = {2025},
  howpublished = {\url{https://ecoinvent.org/database/}},
  note         = {Accessed: 2025-10-17}
}

@misc{carbonminds_database,
  author       = {{Carbon Minds GmbH}},
  title        = {cm.chemicals Database: Life Cycle Inventory Data for Chemicals and Plastics},
  year         = {2025},
  howpublished = {\url{https://www.carbon-minds.com/products/data/carbon-footprint-and-lca-data/}},
  note         = {Accessed: 2025-10-17}
}

@techreport{stellner2023_carbonminds,
  author       = {Stellner, Laura and Kalousdian, Aline and K{\"a}telh{\"o}n, Arne and V{\"o}gler, Oskar and Hermanns, Ronja and Suh, Sangwon and Bardow, Andr{\'e} and Meys, Raoul},
  title        = {Methodology cm.chemicals},
  institution  = {Carbon Minds GmbH},
  address      = {Cologne, Germany},
  year         = {2023},
  version      = {2.00},
  url          = {https://www.carbon-minds.com/wp-content/uploads/2024/01/cmchemicals-methodology.pdf},
  note         = {Version 2.00, July 2023}
}

@article{huber2022approach,
  author       = {Huber, Elena and Bach, Vanessa and Holzapfel, Peter and Blizniukova, Daria and Finkbeiner, Matthias},
  title        = {An Approach to Determine Missing Life Cycle Inventory Data for Chemicals {(RREM)}},
  journal      = {Sustainability},
  year         = {2022},
  volume       = {14},
  number       = {6},
  pages        = {3161},
  publisher    = {MDPI},
  doi          = {10.3390/su14063161},
  url          = {https://doi.org/10.3390/su14063161}
}

@article{parvatker2019cradle,
  author       = {Parvatker, Abhijeet G. and Tunceroglu, Huseyin and Sherman, Jodi D. and Coish, Philip and Anastas, Paul and Zimmerman, Julie B. and Eckelman, Matthew J.},
  title        = {Cradle-to-Gate Greenhouse Gas Emissions for Twenty Anesthetic Active Pharmaceutical Ingredients Based on Process Scale-Up and Process Design Calculations},
  journal      = {ACS Sustainable Chemistry \& Engineering},
  year         = {2019},
  volume       = {7},
  number       = {7},
  pages        = {6580--6591},
  publisher    = {American Chemical Society},
  doi          = {10.1021/acssuschemeng.8b05473},
  url          = {https://doi.org/10.1021/acssuschemeng.8b05473}
}

@techreport{dysert2016aace,
  author       = {Dysert, L. R. and Christesen, P.},
  title        = {{AACE International Recommended Practice No. 18R-97: Cost Estimate Classification System -- As Applied in Engineering, Procurement, and Construction for the Process Industries}},
  institution  = {AACE International Inc.},
  year         = {2016},
  address      = {New York, USA},
  type         = {Recommended Practice},
  url          = {https://services.austintexas.gov/edims/document.cfm?id=280770},
  note         = {Accessed: 17 October 2025}
}

@techreport{doka2013updates,
  author       = {Doka, Gabor},
  title        = {Updates to Life Cycle Inventories of Waste Treatment Services -- {Part II}: Waste Incineration},
  institution  = {Doka Life Cycle Assessments, Zurich},
  address      = {Zurich, Switzerland},
  year         = {2013},
  type         = {Technical Report},
  url          = {http://www.lcaforum.ch/inventories/Hintergrund/Doka_2013-ecoinvent_MSWI_updateLCI.pdf},
  note         = {Accessed: 2025-10-17}
}

@techreport{doka2003life,
  author       = {Doka, Gabor},
  title        = {Life Cycle Inventories of Waste Treatment Services},
  institution  = {Ecoinvent},
  journal      = {Final report ecoinvent},
  volume       = {13},
  address      = {Zurich, Switzerland},
  year         = {2003},
  type         = {Technical Report},
  note         = {Accessed: 2025-10-17},
  url          = {https://www.doka.ch/13_I_WasteTreatmentGeneral.pdf}
}

@book{heijungs2024probability,
  author       = {Heijungs, Reinout},
  title        = {Probability, Statistics and Life Cycle Assessment: Guidance for Dealing with Uncertainty and Sensitivity},
  year         = {2024},
  publisher    = {Springer Nature},
  address      = {Cham, Switzerland},
  edition      = {1st},
  doi          = {10.1007/978-3-031-49317-1},
  url          = {https://doi.org/10.1007/978-3-031-49317-1}
}

@techreport{althaus2007life,
  author       = {Althaus, H.-J. and Chudacoff, Mike and Hischier, Roland and Jungbluth, Niels and Osses, Maggie and Primas, Alex and Hellweg, Stefanie},
  title        = {Life Cycle Inventories of Chemicals},
  institution  = {Swiss Centre for Life Cycle Inventories},
  address      = {D{\"u}bendorf, Switzerland},
  year         = {2007},
  type         = {Final Report, ecoinvent data v2.0, No. 8},
  note         = {Accessed: 17 October 2025}
}

@article{jacob2018statistics,
  author       = {Jacob, Philipp-Maximilian and Lapkin, Alexei},
  title        = {Statistics of the Network of Organic Chemistry},
  journal      = {Reaction Chemistry \& Engineering},
  year         = {2018},
  volume       = {3},
  number       = {1},
  pages        = {102--118},
  publisher    = {Royal Society of Chemistry},
  doi          = {10.1039/C7RE00129K},
  url          = {https://doi.org/10.1039/C7RE00129K}
}

@techreport{ICCA_Oxford2019,
  author       = {{International Council of Chemical Associations (ICCA)} and {Oxford Economics}},
  title        = {The Global Chemical Industry: Catalyzing Growth and Addressing Our World’s Sustainability Challenges},
  institution  = {ICCA / Oxford Economics},
  year         = {2019},
  url          = {https://icca-chem.org/wp-content/uploads/2020/10/Catalyzing-Growth-and-Addressing-Our-Worlds-Sustainability-Challenges-Report.pdf},
  lastaccessed = {2025-10-17}
}

@misc{JRC,
  author       = {{European Commission, Joint Research Centre}},
  title        = {Safe and Sustainable by Design},
  year         = {2023},
  howpublished = {\url{https://research-and-innovation.ec.europa.eu/research-area/industrial-research-and-innovation/chemicals-and-advanced-materials/safe-and-sustainable-design_en}},
  note         = {Accessed: 2025-10-17}
}

@techreport{international2006environmental,
  author       = {{International Organization for Standardization (ISO)}},
  title        = {Environmental Management: Life Cycle Assessment -- Principles and Framework},
  institution  = {ISO},
  number       = {ISO 14040:2006},
  year         = {2006},
  address      = {Geneva, Switzerland},
  type         = {International Standard}
}

@techreport{van2019safe,
  author       = {van der Waals, Jochem and
                  Falk, Andreas and
                  Fantke, Peter and
                  Filippousi, Paraskevi and
                  Flipphi, Ronald and
                  Mottet, Denis and
                  Trier, Xenia},
  title        = {Safe-by-design for materials and chemicals: Towards an innovation programme in Horizon Europe},
  institution  = {Ministry of Infrastructure and Water Management, the Netherlands},
  year         = {2019},
  type         = {Report},
  doi          = {10.5281/zenodo.3254382},
  url          = {https://doi.org/10.5281/zenodo.3254382},
  note         = {Accessed: 2025-10-17}
}

@misc{REACH_2025,
  author       = {{European Chemicals Agency (ECHA)}},
  title        = {{REACH -- Registration, Evaluation, Authorisation and Restriction of Chemicals Regulation: Information on Registered Substances}},
  year         = {2025},
  howpublished = {\url{https://echa.europa.eu/information-on-chemicals/registered-substances}},
  note         = {Accessed: 12 August 2025}
}

@article{zhang2024enhanced,
  author       = {Zhang, Dachuan and Wang, Zhanyun and Oberschelp, Christopher and Bradford, Eric and Hellweg, Stefanie},
  title        = {Enhanced Deep-Learning Model for Carbon Footprints of Chemicals},
  journal      = {ACS Sustainable Chemistry \& Engineering},
  year         = {2024},
  volume       = {12},
  number       = {7},
  pages        = {2700--2708},
  publisher    = {American Chemical Society},
  doi          = {10.1021/acssuschemeng.3c07038},
  url          = {https://doi.org/10.1021/acssuschemeng.3c07038}
}

@article{zuiderveen2023potential,
  author       = {Zuiderveen, Emma A. R. and Kuipers, Koen J. J. and Caldeira, Carla and Hanssen, Steef V. and Van Der Hulst, Mitchell K. and De Jonge, Melinda M. J. and Vlysidis, Anestis and Van Zelm, Rosalie and Sala, Serenella and Huijbregts, Mark A. J.},
  title        = {The potential of emerging bio-based products to reduce environmental impacts},
  journal      = {Nature Communications},
  volume       = {14},
  number       = {1},
  pages        = {8521},
  year         = {2023},
  publisher    = {Nature Publishing Group},
  doi          = {10.1038/s41467-023-43797-9},
  url          = {https://doi.org/10.1038/s41467-023-43797-9}
}

@article{lee2019comprehensive,
  author    = {Lee, Sang Yup and Kim, Hyun Uk and Chae, Tong Un and Cho, Jae Sung and Kim, Je Woong and Shin, Jae Ho and Kim, Dong In and Ko, Yoo-Sung and Jang, Woo Dae and Jang, Yu-Sin},
  title     = {A comprehensive metabolic map for production of bio-based chemicals},
  journal   = {Nature Catalysis},
  volume    = {2},
  number    = {1},
  pages     = {18--33},
  year      = {2019},
  doi       = {10.1038/s41929-018-0212-4},
  url       = {https://doi.org/10.1038/s41929-018-0212-4},
  publisher = {Nature Publishing Group}
}

@article{rahimi2017chemical,
  author    = {Rahimi, AliReza and Garc{\'\i}a, Jeannette M.},
  title     = {Chemical recycling of waste plastics for new materials production},
  journal   = {Nature Reviews Chemistry},
  volume    = {1},
  number    = {6},
  article   = {0046},
  year      = {2017},
  publisher = {Nature Publishing Group},
  doi       = {10.1038/s41570-017-0046},
  url       = {https://doi.org/10.1038/s41570-017-0046}
}

@inproceedings{satopaa2011finding,
  author       = {Satopaa, Ville and Albrecht, Jeannie and Irwin, David and Raghavan, Barath},
  title        = {Finding a ``Kneedle'' in a Haystack: Detecting Knee Points in System Behavior},
  booktitle    = {2011 31st International Conference on Distributed Computing Systems Workshops {(ICDCSW)}},
  pages        = {166--171},
  year         = {2011},
  organization = {IEEE},
  doi          = {10.1109/ICDCSW.2011.20},
  url          = {https://ieeexplore.ieee.org/document/5961514},
  note         = {Accessed: 2025-10-17}
}

@techreport{UNEP_ICCA_knowledge_sharing,
  author       = {{UN Environment} and {International Council of Chemical Associations (ICCA)}},
  title        = {Knowledge Management and Information Sharing for the Sound Management of Industrial Chemicals},
  institution  = {United Nations Environment Programme (UNEP) / ICCA},
  year         = {2015},
  url          = {https://www.saicm.org/Portals/12/Documents/EPI/Knowledge_Information_Sharing_Study_UNEP_ICCA.pdf},
  note         = {Accessed: 2025-10-17}
}

@misc{OECD_HPV_chemicals,
  author       = {{Organisation for Economic Co-operation and Development (OECD)}},
  title        = {{OECD} Existing Chemicals Database},
  year         = {2025},
  howpublished = {\url{https://hpvchemicals.oecd.org/ui/AllChemicals.aspx}},
  note         = {Accessed: 17 October 2025}
}

@article{Gao2026GNN,
  author    = {Gao, Qinghe and Schulze Balhorn, Lukas and Laera, Alessandro and Meys, Raoul and Go{\ss}en, Jonas and Weber, Jana M. and Wernet, Gregor and Schweidtmann, Artur M.},
  title     = {Environmental impacts prediction using graph neural networks on molecular graphs},
  journal   = {Computers \& Chemical Engineering},
  year      = {2026},
  volume    = {204},
  pages     = {109362},
  doi       = {10.1016/j.compchemeng.2025.109362},
  publisher = {Elsevier},
  url       = {https://doi.org/10.1016/j.compchemeng.2025.109362}
}

@book{secretariat1997technical,
  author       = {{Secretariat of the Basel Convention on the Control of Transboundary Movements of Hazardous Wastes and Their Disposal}},
  title        = {Technical Guidelines on Incineration on Land},
  year         = {1997},
  publisher    = {RSM Press},
  address      = {Geneva, Switzerland},
  url          = {http://www.basel.int/Portals/4/download.aspx?d=UNEP-CHW-WAST-GUID-IncinerationLand.English.pdf},
  note         = {Accessed: 2025-10-24}
}

@techreport{ozone_depletion_report,
  author       = {{Union of International Associations}},
  title        = {Montreal Protocol on Substances that Deplete the Ozone Layer},
  institution  = {Union of International Associations},
  year         = {1987},
  url          = {https://uia.org/s/or/en/1100031640},
  note         = {Accessed: 2025-10-24}
}

@article{astrup2005chromium,
  author    = {Astrup, Thomas and Rosenblad, C. and Trapp, Stefan and Christensen, Thomas H{\o}jlund},
  title     = {Chromium Release from Waste Incineration Air-Pollution-Control Residues},
  journal   = {Environmental Science \& Technology},
  year      = {2005},
  volume    = {39},
  number    = {9},
  pages     = {3321--3329},
  publisher = {American Chemical Society},
  doi       = {10.1021/es049346q},
  url       = {https://doi.org/10.1021/es049346q}
}

@misc{EuroEnvironmental2024,
  author       = {{Euro Environmental Ltd.}},
  title        = {{Case Study: Managing Chromium VI Exposure in Waste Recycling Incinerators}},
  year         = {2024},
  month        = {Jul},
  url          = {https://www.euroenvironmental.co.uk/news/item/managing-chromium-vi-exposure-in-waste-recycling-incinerators},
  note         = {Accessed: 2025-10-31}
}

@article{chen2025semanet,
  title        = {{SemaNet: Bridging Words and Numbers for Predicting Missing Environmental Data in Life Cycle Assessment}},
  author       = {Chen, Bin and Chen, Hong and Quan, Zhishan and He, Wei and Kadirkamanathan, Visakan and Casamayor, Jose L. and Xing, Wei W.},
  journal      = {Environmental Science \& Technology},
  year         = {2025},
  volume       = {59},
  number       = {39},
  pages        = {21131--21146},
  doi          = {10.1021/acs.est.5c07557},
  url          = {https://doi.org/10.1021/acs.est.5c07557},
  publisher    = {American Chemical Society}
}

@article{rada2021regulatory,
  author    = {Rada, Elena Cristina and Schiavon, Marco and Torretta, Vincenzo},
  title     = {A Regulatory Strategy for the Emission Control of Hexavalent Chromium from Waste-to-Energy Plants},
  journal   = {Journal of Cleaner Production},
  volume    = {278},
  pages     = {123415},
  year      = {2021},
  doi       = {10.1016/j.jclepro.2020.123415},
  url       = {https://doi.org/10.1016/j.jclepro.2020.123415},
  publisher = {Elsevier}
}

@book{ATSDR2012Chromium,
  author       = {Wilbur, Sharon B and Abadin, Henry and Fay, Mike and Tencza, Brian and Yu, Dianyi and Ingerman, Lisa and Klotzbach, Julie and James, Shelly},
  title        = {Toxicological Profile for Chromium},
  year         = {2012},
  publisher    = {U.S. Department of Health and Human Services, Public Health Service},
  address      = {Atlanta, GA},
  institution  = {Agency for Toxic Substances and Disease Registry (ATSDR)},
  url          = {https://www.ncbi.nlm.nih.gov/books/NBK158858/},
  note         = {Accessed: 2025-11-11}
}

@article{blasenbauer2020legal,
  title        = {Legal situation and current practice of waste incineration bottom ash utilisation in Europe},
  author       = {Blasenbauer, Dominik and Huber, Florian and Lederer, Jakob and Quina, Margarida J and Blanc-Biscarat, Denise and Bogush, Anna and Bontempi, Elza and Blondeau, Julien and Chimenos, Josep Maria and Dahlbo, Helena and others},
  journal      = {Waste Management},
  volume       = {102},
  pages        = {868--883},
  year         = {2020},
  doi          = {10.1016/j.wasman.2019.11.031},
  url          = {https://doi.org/10.1016/j.wasman.2019.11.031},
  publisher    = {Elsevier}
}

@article{meys2021achieving,
  author       = {Meys, Raoul and Kätelhöhn, Arne and Bachmann, Marvin and Winter, Benedikt and Zibunas, Christian and Suh, Sangwon and Bardow, André},
  title        = {Achieving net-zero greenhouse gas emission plastics by a circular carbon economy},
  journal      = {Science},
  volume       = {374},
  number       = {6563},
  pages        = {71--76},
  year         = {2021},
  publisher    = {American Association for the Advancement of Science},
  doi          = {10.1126/science.abg9853},
  url          = {https://doi.org/10.1126/science.abg9853},
}

@misc{BaselConvention2025,
  title        = {Basel Convention on the Control of Transboundary Movements of Hazardous Wastes and Their Disposal: Text and Annexes (revised 2025)},
  author       = {{Basel Convention Secretariat, United Nations Environment Programme}},
  year         = {2025},
  month        = {July},  
  howpublished = {United Nations Environment Programme, Secretariat of the Basel Convention},  
  address      = {Geneva, Switzerland},  
  url          = {https://www.basel.int/Portals/4/download.aspx?e=UNEP-CHW-IMPL-CONVTEXT-2025.English.pdf}
}

@article{bustillo2023rise,
  title     = {The rise of automated curiosity-driven discoveries in chemistry},
  author    = {Bustillo, Latimah and Laino, Teodoro and Rodrigues, Tiago},
  journal   = {Chemical Science},
  volume    = {14},
  number    = {38},
  pages     = {10378--10384},
  year      = {2023},
  doi       = {10.1039/D3SC03367H},
  url       = {https://doi.org/10.1039/D3SC03367H},
  publisher = {Royal Society of Chemistry}
}

@article{Greed2022TurningSolvents,
  author  = {Greed, Sam},
  title   = {Turning solvents into something},
  journal = {Nature Reviews Chemistry},
  year    = {2022},
  volume  = {6},
  pages   = {519},
  doi     = {10.1038/s41570-022-00414-5},
  url     = {https://www.nature.com/articles/s41570-022-00414-5}
}

@misc{CDR2020,
  title        = {2020 Chemical Data Reporting (CDR) Data},
  author       = {{U.S. Environmental Protection Agency}},
  organization = {U.S. Environmental Protection Agency, Office of Pollution Prevention and Toxics (OPPT)},
  year         = {2020},
  note         = {Accessed: 2025-12-22},
  url          = {https://www.epa.gov/chemical-data-reporting/access-chemical-data-reporting-data},
}

@misc{SPIN2025,
  title        = {{SPIN: Substances in Preparations in Nordic Countries}},
  author       = {{Nordic Council of Ministers}},
  organization = {Nordic Council of Ministers},
  year         = {2025},
  note         = {Accessed: 2025-12-22},
  url          = {http://spin2000.net/},
}

@misc{Meti2025,
  title        = {Annual Manufacture and Import Quantities of General Chemical Substances in Japan},
  author       = {{Ministry of Economy, Trade and Industry (METI), Japan}},
  organization = {Ministry of Economy, Trade and Industry, Chemical Management Division},
  year         = {2025},
  note         = {Accessed: 2025-12-22},
  url          = {https://www.meti.go.jp/policy/chemical_management/kasinhou/information/volume_general.html},
}

@article{life_cycle_assessment_chemical_industry_review,
  author       = {Santos, Andreia and Barbosa-P{\'o}voa, Ana and Carvalho, Ana},
  title        = {Life Cycle Assessment in Chemical Industry – A Review},
  journal      = {Current Opinion in Chemical Engineering},
  year         = {2019},
  volume       = {26},
  number       = {},
  pages        = {139--147},
  publisher    = {Elsevier},
  doi          = {10.1016/j.coche.2019.09.009},
  url          = {https://doi.org/10.1016/j.coche.2019.09.009}
}
